\PassOptionsToPackage{section}{placeins}
\documentclass[]{relaxed_system_lab}
\usepackage[utf8]{inputenc}
\usepackage[T1]{fontenc}
\usepackage{charter}
\usepackage{amsmath,amssymb,mathtools}
\usepackage{xspace,array,seqsplit,enumitem,tabularx}
\usepackage{needspace}
\usepackage{float}
\usepackage{wrapfig}
\usepackage{xurl}
\usepackage{mdframed}
\mdfsetup{nobreak=true}
\usepackage{algorithm}
\usepackage{algpseudocode}

\newcommand{\sys}{{\upshape\textsc{Afore}}\xspace}

\providecommand{\Description}[2][]{}

\mdfdefinestyle{insightbox}{
  backgroundcolor=green!5!white,linecolor=green!40!black,
  linewidth=0.3pt,roundcorner=2pt,
  innerleftmargin=4pt,innerrightmargin=4pt,
  innertopmargin=3pt,innerbottommargin=3pt}
\mdfdefinestyle{motivationbox}{
  backgroundcolor=blue!5!white,linecolor=blue!40!black,
  linewidth=0.3pt,roundcorner=2pt,
  innerleftmargin=4pt,innerrightmargin=4pt,
  innertopmargin=3pt,innerbottommargin=3pt}
\mdfdefinestyle{codebox}{
  backgroundcolor=black!4!white,linecolor=black!40!black,
  linewidth=0.3pt,roundcorner=2pt,
  innerleftmargin=4pt,innerrightmargin=4pt,
  innertopmargin=3pt,innerbottommargin=3pt}
\title{\sys: Attention--FFN Disaggregation with Overlapped Reconfiguration of Experts}
\hypersetup{pdftitle={AFORE: Attention--FFN Disaggregation with Overlapped Reconfiguration of Experts}}

\renewcommand{\authorlist}{%
  \authorformat[1]{Wenshuang Li},
  \authorformat[2]{Youhe Jiang},
  \authorformat[1]{You Peng}\\[0.5mm]
  \authorformat[3]{Jiawei Jiang},
  \authorformat[1]{Binhang Yuan}%
}
\renewcommand{\affiliationlist}{%
  \affiliationformat[1]{The Hong Kong University of Science and Technology}\\[0.5mm]
  \affiliationformat[2]{University of Cambridge}\qquad
  \affiliationformat[3]{Wuhan University}%
}
\hypersetup{pdfauthor={Wenshuang Li, Youhe Jiang, You Peng, Jiawei Jiang, Binhang Yuan}}
\abstract{Mixture-of-Experts (MoE) models are increasingly leveraged in deployed large-scale LLM services, but efficient MoE serving remains challenging due to large expert parameters, input-dependent expert activation, and dynamic workloads. Expert parallelism is commonly used to distribute expert computation across GPUs, while attention-FFN disaggregation (AFD) further improves serving efficiency by separating attention and feed-forward computation into independent worker pools. However, we observe that a naive AFD implementation could make expert load imbalance more harmful: once FFN computation becomes an independent pipeline stage, overloaded experts directly slow the FFN stage and degrade end-to-end serving performance. 

To solve this problem, we present \sys, a timely expert reconfiguration system for AFD-based MoE serving. \sys exploits two architectural properties of AFD. First, AFD exposes the expert-token distribution of upcoming microbatches before they reach FFN execution, enabling placement decisions based on near-future demand instead of stale historical profiles. Second, AFD creates a pipeline window in which expert migration for a target microbatch can be overlapped with the computation of preceding in-flight microbatches. Based on these insights, \sys formulates expert reconfiguration as a microbatch-aware scheduling problem and uses a migration-aware scheduler to decide when and which experts to migrate. \sys further implements lightweight demand prefetching and NVLink-based GPU-GPU expert migration to reduce reconfiguration overhead. Evaluation on a 110B-parameter MoE model across four dynamic workloads shows that \sys improves output throughput by \textbf{10.1--17.6\%} and reduces P95 inter-token latency by \textbf{7.1--9.5\%} compared with the strongest competing baseline. Compared with static placement, \sys improves throughput by \textbf{29.8\%} on average and reduces P95 inter-token latency by \textbf{18.2\%} on average. Migration profiling further shows that AFD pipeline overlap can fully hide expert-migration latency.
}

\begin{document}
\maketitle

\section{Introduction}
\label{sec:intro}

Mixture-of-Experts (MoE) models are increasingly used by deployed large-scale large language model (LLM) services, including interactive assistants, coding copilots, search, data analysis, enterprise automation, and agentic applications~\cite{fedus2022switch,jiang2024mixtral,du2022glam,liu2024deepseek,dai2024deepseekmoe,yang2025qwen3}. By activating only a small subset of experts for each token, MoE models increase model capacity while keeping per-token computation manageable~\cite{lepikhin2020gshard,fedus2022switch}. However, serving MoE inference at scale remains costly: expert parameters are large, expert activation is input-dependent, and workloads change over time~\cite{rajbhandari2022deepspeed,he2021fastmoe,frantar2023qmoe}, where the existing solutions are not sufficient. For example, expert parallelism (EP) distributes experts across GPUs to scale expert execution~\cite{rajbhandari2022deepspeed}, and attention-FFN disaggregation (AFD)~\cite{zhu2025megascale} further improves resource utilization by separating attention and feed-forward computation into distinct worker pools. Yet in AFD, skewed expert routing directly bottlenecks the independent FFN stage, creates pipeline backpressure, and degrades end-to-end serving performance. In this paper, we explore \textit{how to reduce dynamic expert-load imbalance in AFD-based MoE serving without incurring reconfiguration overhead that erodes the resulting performance gains.}

Addressing this problem is essential because the efficiency of MoE serving increasingly determines the cost and responsiveness of production LLM services~\cite{he2021fastmoe,hwang2023tutel,rajbhandari2022deepspeed,go2025moetuner}. As modern MoE models grow and workloads become more heterogeneous, small inefficiencies in expert execution can translate into substantial GPU underutilization, reduced output throughput, and higher tail latency~\cite{jiang2024mixtral,liu2024deepseek,yang2025qwen3,hwang2023tutel,go2025moetuner,cong2024prediction}. AFD makes this issue especially consequential: once FFN computation is isolated as a separate pipeline stage, imbalance among expert workers can throttle the entire serving pipeline rather than remaining a localized execution inefficiency~\cite{zhu2025megascale,chen2024efficient,wang2025step,song2026theoretically}. Addressing FFN-side expert imbalance is therefore essential for realizing the scalability benefits of AFD while preserving the latency guarantees required by interactive LLM applications~\cite{zhu2025megascale,chen2024efficient,wang2025step,hwang2023tutel,go2025moetuner}.

\begin{figure}[t]
    \centering
    \includegraphics[trim=0bp 8.70bp 0bp 0bp,clip,width=4.5in]{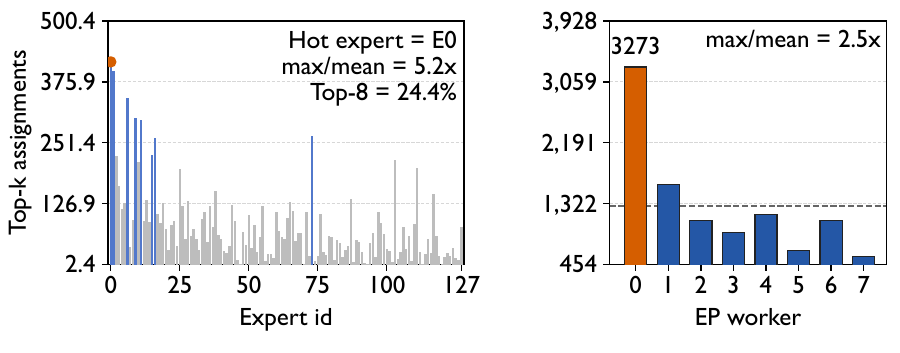}
    \caption{\textbf{\underline{Left:}} Expert routing load in a representative AFD window. With top-$k=8$, a small set of hot experts receives a disproportionate fraction of routed tokens. \textbf{\underline{Right:}} The resulting load across eight EP workers under static expert placement, where the most loaded worker reaches 2.5$\times$ the average load.}
    \Description{For a representative layer-3 ShareGPT microbatch, the left panel shows top-k assignment counts across 128 experts and highlights the eight most loaded experts. The right panel aggregates those assignments over eight statically placed EP workers; worker 0 receives 3273 assignments and reaches 2.5 times the mean worker load.}
    \label{fig:intro-moe-imbalance}
\end{figure}

\begin{figure}[t]
    \centering
    \includegraphics[trim=0bp 8.70bp 0bp 0bp,clip,width=4.5in]{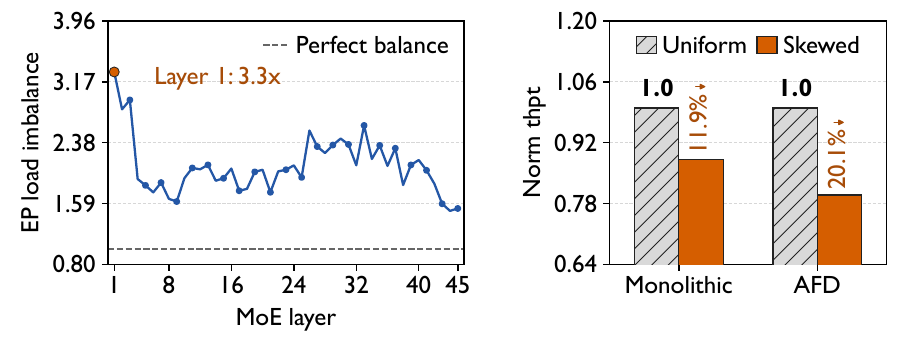}
    \caption{\textbf{\underline{Left:}} EP load imbalance across MoE layers on ShareGPT, with the most imbalanced layer reaching 3.3$\times$ imbalance of the average worker load. \textbf{\underline{Right:}} Normalized throughput under synthetic uniform and skewed routing. Skewed routing reduces throughput by 11.9\% under monolithic attention-FFN execution and by 20.1\% under AFD.}
    \Description{The left panel plots each MoE layer's 99th-percentile maximum-to-mean EP worker load ratio over ShareGPT steps 612 through 868 and all three microbatch positions; layer 1 peaks at 3.3 times. The right panel compares normalized throughput under synthetic uniform and half-hot routing, showing reductions of 11.9 percent for monolithic execution and 20.1 percent for AFD.}
    \label{fig:intro-imbalance-effects}
    \label{fig:intro-layerwise-imbalance}
    \label{fig:intro-afd-imbalance-impact}
\end{figure}

The key system obstacle is that the existing techniques for scaling MoE serving improve capacity and resource utilization but \textit{do not remove the dynamic imbalance that arises from input-dependent expert routing}. \underline{First}, \textit{EP} partitions expert parameters and computation across GPUs, enabling large expert layers to exceed the memory and compute capacity of a single device~\cite{rajbhandari2022deepspeed,he2021fastmoe,go2025moetuner,hwang2023tutel}; however, such an approach also makes expert placement and resource allocation central to performance, because GPUs hosting popular experts become stragglers while others remain underutilized. This imbalance is difficult to avoid statically: different requests, layers, and generation phases can activate substantially different experts, producing the skewed expert-load distributions shown in~\autoref{fig:intro-moe-imbalance} and across-layer imbalance shown in the left panel of~\autoref{fig:intro-imbalance-effects}.

\noindent\underline{Second}, \textit{AFD} provides a complementary architecture by separating attention and FFN computation into independently provisioned worker pools~\cite{zhu2025megascale,song2026theoretically,wang2025step,chen2024efficient}, but this separation makes the FFN/expert stage a distinct pipeline bottleneck rather than merely one component of monolithic execution. To examine the performance impact of routing skew, we compare synthetic uniform and skewed routing under monolithic attention-FFN execution and AFD. As shown in the right panel of~\autoref{fig:intro-imbalance-effects}, skewed routing reduces normalized throughput by 11.9\% under monolithic execution and by 20.1\% under AFD. Thus, existing scalability mechanisms address model capacity and coarse resource provisioning, but they are insufficient for handling rapidly changing expert-load skew: an overloaded FFN worker in AFD slows the independent FFN stage, creates pipeline bubbles, backpressures the attention stage, and degrades end-to-end serving performance. 

A natural approach to mitigating expert load imbalance is by reconfiguration, which adjusts expert placement according to runtime load~\cite{go2025moetuner,wu2025accelerating,singh2025elasticmoe}. 
Reconfiguration can improve expert utilization and reduce stragglers, but existing reconfiguration mechanisms face two fundamental challenges:
\begin{itemize}[leftmargin=*]
    \item \textbf{\underline{C1:} Dynamic and unpredictable demand.} Expert demand changes over time because token routing depends on the input workload, model layer, sequence position, and request mix~\cite{cong2024prediction,hwang2023tutel}. Static placement, historical statistics, or coarse-grained profiling can capture only average or past expert popularity; they cannot reliably describe the expert-token distribution of upcoming execution~\cite{go2025moetuner,nvidia2025eplb}. As a result, reconfiguration decisions may react too late or optimize for stale load patterns.

    \vspace{0.5em}
    \item \textbf{\underline{C2:} Non-negligible system overhead.} Expert reconfiguration introduces runtime overhead from placement scheduling, expert-weight migration, routing-metadata updates, and worker coordination~\cite{singh2025elasticmoe,wu2025accelerating}. If these operations are exposed on the critical path, the cost of reconfiguration can offset its load-balancing benefit and even degrade overall performance~\cite{balmau2025accelerating,vllm2025swiftbalancer}.
\end{itemize}

Our key observation is that AFD, while making expert imbalance more harmful, also provides two architectural opportunities for addressing these challenges. \textit{In particular, AFD both amplifies expert imbalance and exposes the temporal structure needed to fix it.}

\begin{itemize}[leftmargin=*]
    \item \textbf{\underline{O1:} Microbatch-level demand is visible.} AFD processes requests by a disaggregated microbatch pipeline, where a target microbatch is handled by the attention side before reaching the FFN/expert side~\cite{zhu2025megascale}. This pipeline structure exposes the expert-token distribution of the target microbatch before its FFN execution. Therefore, reconfiguration decisions can use near-future demand already available in the pipeline, enabling timely expert placement adaptation under dynamic workloads (addressing \textbf{\underline{C1}}).

    \vspace{0.5em}
    \item \textbf{\underline{O2:} reconfiguration overhead can be overlapped.} AFD creates a pipeline window between observing the target microbatch's expert demand and executing that microbatch on the FFN/expert side~\cite{zhu2025megascale,song2026theoretically}. During this window, preceding in-flight microbatches continue executing on FFN workers. This pipeline structure allows expert migration for the target microbatch to overlap with preceding microbatch computation, reducing the migration overhead exposed on the critical path (addressing \textbf{\underline{C2}}).
\end{itemize}

Based on these insights, we present \sys to accelerate MoE inference serving by \underline{\textbf{a}}ttention--\underline{\textbf{F}}FN
disaggregation with \underline{\textbf{o}}verlapped \underline{\textbf{r}}econfiguration of \underline{\textbf{e}}xperts. \sys adapts expert placement over time using microbatch-level expert-demand information and overlaps expert migration with the computation of preceding in-flight microbatches. At its core, \sys treats expert reconfiguration as a microbatch-aware scheduling problem: for each target microbatch, the system estimates the FFN-stage imbalance under the current placement, evaluates candidate expert-placement changes, and triggers migration only when the expected reduction in FFN bottlenecks outweighs the migration overhead exposed on the critical path. This design improves FFN-stage balance and end-to-end serving efficiency under dynamic workloads.
Our contributions are summarized as follows:

\vspace{0.5em}
\textbf{\underline{Contribution 1:}} We formulate temporal expert reconfiguration in AFD-based MoE serving as an online expert-placement scheduling problem. The problem jointly determines when to adjust expert placement and which experts to migrate for upcoming microbatches, while accounting for both near-future expert demand and exposed migration overhead. To solve this problem efficiently, we design a migration-aware scheduling algorithm that uses prefetched expert-token distributions to estimate FFN-stage imbalance, evaluates the benefit of candidate placement changes, and performs reconfiguration only when it reduces the critical-path cost of FFN execution.

\vspace{0.5em}
\textbf{\underline{Contribution 2:}} We design and implement \sys with runtime support for microbatch-level demand prefetching and efficient expert migration. \sys introduces a lightweight prefetching interface that collects the expert-token distribution of target microbatches before they reach FFN execution, enabling placement decisions based on near-future demand. \sys overlaps expert migration with the computation of preceding in-flight microbatches and applies placement updates at microbatch boundaries to avoid disrupting ongoing execution. To reduce migration time, \sys uses NVLink-based GPU-GPU expert transfer for fast expert reconfiguration.

\vspace{0.5em}
\textbf{\underline{Contribution 3:}} We evaluate \sys on a 110B-parameter MoE model (i.e., GLM-4.5-Air) under four dynamic serving workloads and compare \sys against static placement, statistics-based reconfiguration, reactive load-balancing, and prediction-based reconfiguration baselines. Experimental results show that \sys improves output throughput by \textbf{10.1--17.6\%} and reduces P95 inter-token latency by \textbf{7.1--9.5\%} compared with the strongest competing baseline. Component-wise ablation shows that \sys improves throughput by \textbf{29.8\%} on average and reduces P95 inter-token latency by \textbf{18.2\%} on average over static placement. Migration profiling further shows that AFD pipeline overlap can fully hide expert-migration latency.

\vspace{0.5em}
The rest of this paper is organized as follows. \S\ref{sec:background} provides background and related work. \S\ref{sec:motivation} presents the motivation for temporal expert reconfiguration in AFD-based MoE serving. \S\ref{sec:overview} introduces the overall architecture of \sys. \S\ref{sec:sched} presents the scheduling framework for expert reconfiguration. \S\ref{sec:systemdesign} describes the system design and runtime implementation. \S\ref{sec:eval} presents the system evaluation. \S\ref{sec:conclusion} concludes the paper.

\section{Background and Related Work}
\label{sec:background}

\noindent\textbf{Expert parallelism for MoE deployment.}
MoE models improve the parameter efficiency of LLMs by activating only a small subset of experts for each token~\cite{lepikhin2020gshard,fedus2022switch,du2022glam,jiang2024mixtral,dai2024deepseekmoe,liu2024deepseek,yang2025qwen3}. This sparse activation enables larger parameter space without a proportional increase in per-token computation, which also makes serving more complex because expert parameters and expert computation must be distributed across devices. Recent MoE serving systems commonly rely on expert parallelism to partition experts across GPUs and dispatch tokens to the corresponding expert workers~\cite{he2021fastmoe,hwang2023tutel,rajbhandari2022deepspeed}. For instance, FastMoE~\cite{he2021fastmoe} and Tutel~\cite{hwang2023tutel} optimize distributed expert execution and token dispatch; DeepSpeed-MoE~\cite{rajbhandari2022deepspeed} provides system support for large-scale MoE training and inference; and recent large MoE models further increase the need for efficient expert-parallel serving~\cite{jiang2024mixtral,liu2024deepseek,yang2025qwen3}. These systems establish expert parallelism as a standard mechanism for scaling MoE deployment. However, expert-parallel performance depends heavily on expert placement and resource allocation, since uneven expert activation could lead to serious stragglers. Our work builds on expert-parallel MoE serving and studies how expert placement should be adapted when the FFN/expert stage is deployed under AFD.

\vspace{0.5em}
\noindent\textbf{Attention-FFN disaggregation.}
A separate thread of effort improves LLM serving by disaggregating computation components with different resource requirements~\cite{zhu2025megascale,wang2025step,chen2024efficient,song2026theoretically}. In AFD, attention and feed-forward computation are placed on separate worker pools so that each side can be provisioned and optimized independently. This design is motivated by the different bottlenecks of the two stages: attention execution is bounded by KV-cache and other relevant memory access, while FFN execution is dominated by dense or expert MLP computation. Recent AFD systems study how to organize the disaggregated pipeline, transfer intermediate activations, and allocate resources between attention and FFN workers~\cite{zhu2025megascale,wang2025step,chen2024efficient}. These works show that AFD is a promising architecture for scalable LLM serving. Our work focuses on maintaining efficient MoE serving under the AFD architecture, where the FFN side may contain expert-parallel computation and its load balance directly affects the service rate of the disaggregated pipeline.

\vspace{0.5em}
\noindent\textbf{MoE workload dynamics and expert-load imbalance.}
Expert-load imbalance is an essential challenge in MoE serving systems because token routing is input-dependent~\cite{lepikhin2020gshard,fedus2022switch,hwang2023tutel,go2025moetuner,cong2024prediction}. Different requests, layers, and generation phases can activate different subsets of experts, causing some experts to receive far more tokens than others. Prior work addresses this issue from several angles~\cite{wu2026easybalance,wang2025amoe}. MoE model designs use auxiliary load-balancing losses or routing constraints to encourage more uniform expert usage~\cite{lepikhin2020gshard,fedus2022switch}; systems such as Tutel optimize token dispatch and communication under skewed activation~\cite{hwang2023tutel}; and expert-placement or prediction methods exploit routing statistics to reduce load skew and communication overhead~\cite{go2025moetuner,cong2024prediction,nvidia2025eplb}. These techniques reduce imbalance, but they generally rely on training-time regularization, static placement, or historical runtime feedback. In contrast, \sys targets serving-time expert-load dynamics under AFD by leveraging the disaggregated microbatch pipeline to make expert-placement decisions with timely workload information.

\vspace{0.5em}
\noindent\textbf{MoE expert reconfiguration.}
Expert reconfiguration dynamically changes expert placement, replication, or resource allocation to adapt to runtime load~\cite{go2025moetuner,wu2025accelerating,singh2025elasticmoe,balmau2025accelerating,vllm2025swiftbalancer,nvidia2025eplb}. For example, placement-oriented systems optimize the mapping between experts and GPUs to reduce load imbalance and communication cost~\cite{go2025moetuner}; elastic MoE systems adjust expert resources under changing workloads~\cite{singh2025elasticmoe,wu2025accelerating}; and online balancing mechanisms migrate or redistribute experts to avoid persistent hot spots~\cite{vllm2025swiftbalancer,nvidia2025eplb}. These approaches demonstrate the value of adapting expert resources at runtime. However, reconfiguration also introduces overhead that can offset the benefit of load balancing and even degrade end-to-end performance. \sys therefore treats reconfiguration not only as a placement problem, but also as a scheduling problem: it overlaps expert migration with the computation of preceding in-flight microbatches in the AFD pipeline, reducing the reconfiguration overhead exposed to serving.

\begin{figure}[t]
    \centering
    \includegraphics[trim=0bp 9.45bp 0bp 0bp,clip,width=4.5in]{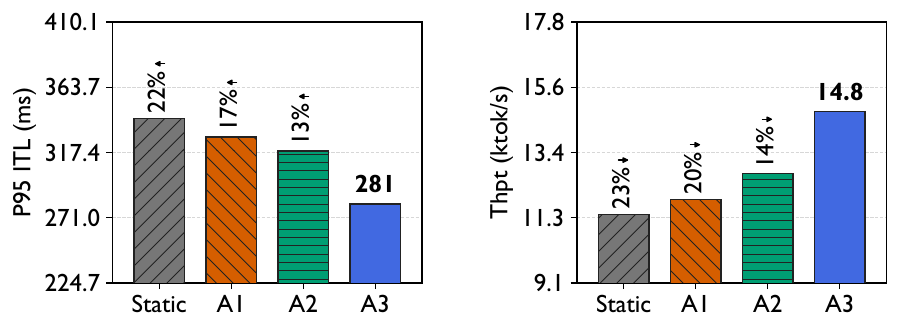}
    \caption{\textbf{\underline{Left:}} P95 inter-token latency under static
    baseline placement and the three progressively enabled reconfiguration approaches.
    \textbf{\underline{Right:}} Corresponding output throughput under the same
    configurations.}
    \Description{The left panel compares P95 inter-token latency and the
    right panel compares output throughput across Static and Approaches 1
    through 3.}
    \label{fig:motivation-results}
\end{figure}

\section{Case Study and Motivation}
\label{sec:motivation}

We first present a controlled case study to understand how expert reconfiguration affects AFD-based MoE serving, where we focus on three key questions: 

\begin{itemize}[leftmargin=*]
    \item \textbf{\underline{Question 1:}} \textit{whether static expert placement is sufficient under dynamic expert demand.}

    \vspace{0.5em}
    \item \textbf{\underline{Question 2:}} \textit{whether microbatch-level demand prefetching can improve reconfiguration decisions.}

    \vspace{0.5em}
    \item \textbf{\underline{Question 3:}} \textit{whether the pipeline execution of AFD can hide expert migration overhead.}
\end{itemize}

\vspace{0.5em}
\noindent\textbf{Experiment setup.}
We serve a 110B-parameter MoE model (\texttt{GLM-4.5-Air}) on a cluster of 16 NVIDIA A100 GPUs under a PD+AFD deployment. The cluster is partitioned into 4 prefill workers,
4 attention/routing workers, and 8 FFN/expert workers; the FFN side uses
expert parallelism degree 8. We replay a dynamic serving workload
in which the expert-token distribution varies across microbatches. All
configurations use the same model, workload, GPU allocation, batching policy,
and FFN expert-parallel layout. We compare four configurations:
(\textbf{\underline{i}}) static expert placement,
(\textbf{\underline{ii}}) expert reconfiguration without prefetch or overlap,
(\textbf{\underline{iii}}) expert reconfiguration with prefetch but without
overlap, and (\textbf{\underline{iv}}) expert reconfiguration with prefetch
and overlap. We report end-to-end throughput and inter-token latency
in~\autoref{fig:motivation-results}.

\vspace{0.5em}
\noindent\textbf{Baseline approach: static expert placement.}
We first evaluate static expert placement, where experts are evenly partitioned
across the 8 FFN workers at initialization, and the placement remains fixed
throughout serving. This configuration introduces no migration overhead, but
it cannot adapt to changes in expert demand dynamics. As shown
in~\autoref{fig:motivation-results}, the baseline reaches
11,357.31\,tokens/s throughput and 341.713\,ms P95 inter-token latency.
Its mean token-assignment EP load imbalance ratio is 1.70$\times$,
causing the FFN stage to wait for overloaded workers.

\begin{figure}[t]
    \centering
    \includegraphics[trim=0bp 9.45bp 0bp 0bp,clip,width=4.5in]{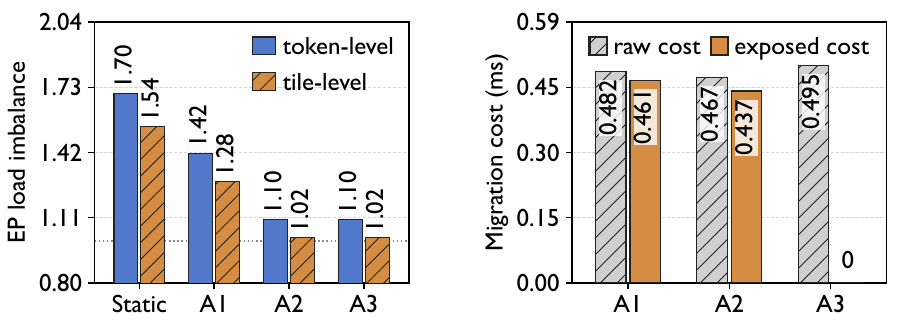}
    \caption{\textbf{\underline{Left:}} EP load imbalance ratio
    under Static and Approaches~1--3 (mean values).
    \textbf{\underline{Right:}} Copy-readiness latency and exposed wait
    under Approaches~1--3 (mean values).}
    \Description{The left panel uses grouped bars to compare mean maximum-to-mean
    EP worker load ratios computed from token assignments and 64-token tiles
    under Static and Approaches 1 through 3. The right uses paired bars to compare mean copy-readiness
    latency and mean exposed wait for Approaches 1 through 3.}
    \label{fig:motivation-design-effects}
    \label{fig:motivation-target-demand}
    \label{fig:motivation-migration-cost}
\end{figure}

\vspace{0.5em}
\noindent\textbf{Approach 1: expert reconfiguration without prefetch or overlap.}
We next evaluate dynamic expert reconfiguration without using demand
prefetching or migration overlap. For a target microbatch $m_t$ in FFN
execution order, let $d_t$ denote its per-expert demand and $d_{h(t)}$
the demand observed at the same layer and microbatch position in the
preceding serving step. This approach uses $d_{h(t)}$ rather than $d_t$
to identify historical hot experts and adjust their placement. When a
reconfiguration decision is made, the system migrates the selected experts and applies the new placement before executing
$m_t$. Compared with Static, this approach lowers the mean token-assignment
EP load imbalance ratio from 1.70$\times$ to 1.42$\times$
(a 16.7\% reduction), improves throughput by 4.5\%, and reduces
P95 inter-token latency by 3.9\%.

\vspace{1em}
\begin{mdframed}[style=motivationbox]
\noindent\textbf{\underline{Insight 1:} dynamic reconfiguration is necessary.}
The improvement over Static shows that expert placement should not remain
fixed under dynamic MoE workloads. Even a simple history-based policy can
reduce FFN-stage stragglers by moving hot experts away from overloaded workers.
This confirms that expert reconfiguration is necessary for maintaining
FFN-stage balance in AFD-based MoE serving.
\end{mdframed}
\vspace{0.5em}

\vspace{0.5em}
\noindent\textbf{Approach 2: expert reconfiguration with prefetch but without overlap.}
We then isolate the benefit of microbatch-level demand prefetching. Instead
of relying on historical demand $d_{h(t)}$, this approach uses the target
microbatch's actual demand $d_t$ to guide expert placement. Compared with
Approach~1, it lowers the mean token-assignment EP load
imbalance ratio from 1.42$\times$ to 1.10$\times$
(a 22.2\% reduction), as shown in the left panel
of~\autoref{fig:motivation-design-effects}. It also improves throughput by 7.1\% and reduces
P95 inter-token latency by 3.0\%.

\vspace{1em}
\begin{mdframed}[style=motivationbox]
\noindent\textbf{\underline{Insight 2:} prefetching improves reconfiguration quality.}
History-based reconfiguration can reduce persistent imbalance, but it may
still adapt to stale load patterns when expert demand changes quickly.
Microbatch-level prefetching exposes the actual expert demand of upcoming
execution, allowing the approach to choose placements that better match
the workload about to enter the FFN stage.
This validates \textbf{\underline{O1}} in \S\ref{sec:intro}.
\end{mdframed}
\vspace{0.5em}

\vspace{0.5em}
\noindent\textbf{Approach 3: expert reconfiguration with prefetch and overlap.}
Finally, we evaluate expert reconfiguration with both demand prefetching
and migration overlap. This approach uses the same prefetched expert-token
distribution as Approach~2, but starts expert migration as soon as the target
microbatch's demand becomes available. The migration overlaps with the FFN-stage
computation of preceding in-flight microbatches; the target uses the new
placement after the required copies complete. Compared with Approach~2,
this approach reduces mean exposed migration wait by 100\%, improves
throughput by 16.4\%, and reduces P95 inter-token latency by
11.8\%. Overall, the complete design improves throughput by
30.3\% and reduces P95 inter-token latency by 17.8\% over Static.
The right panel of~\autoref{fig:motivation-migration-cost} compares mean copy-readiness latency
with mean exposed wait. The mean exposed waits for Approaches~1--3 are
0.461, 0.437, and 0.000\,ms, respectively.

\vspace{1em}
\begin{mdframed}[style=motivationbox,nobreak=true]
\noindent\textbf{\underline{Insight 3:} pipeline overlap reduces reconfiguration cost.}
Prefetching improves the quality of placement decisions, but migration
overhead can compromise the benefit of reconfiguration if it is fully
exposed on the critical path. AFD provides a pipeline window between
observing a target microbatch's expert demand and executing that
microbatch on the FFN side. By overlapping expert migration with the
computation of preceding in-flight microbatches, the system can hide
some migration overhead and improve overall utilization. This validates
\textbf{\underline{O2}} in \S\ref{sec:intro}.
\end{mdframed}
\vspace{0.5em}

\vspace{0.5em}
\noindent\textbf{Implications for system design.}
The case study leads to three design implications. \underline{First}, dynamic
expert reconfiguration is necessary because static expert placement cannot
maintain FFN-stage balance under changing expert demand. \underline{Second},
AFD's microbatch-level demand prefetching benefits reconfiguration by
exposing the exact near-future expert demand before FFN execution.
\underline{Third}, AFD's microbatch pipeline enables expert migration to
overlap with preceding microbatch computation, further improving system
utilization by reducing exposed migration overhead. We design \sys based on
these implications: \sys formulates expert placement as a microbatch-aware
scheduling problem, uses demand prefetching to guide reconfiguration,
and schedules fast expert migration within the AFD pipeline window.

\begin{figure}[t]
    \centering
    \includegraphics[trim=0bp 1.28bp 0bp 0bp,clip,width=6.0in]{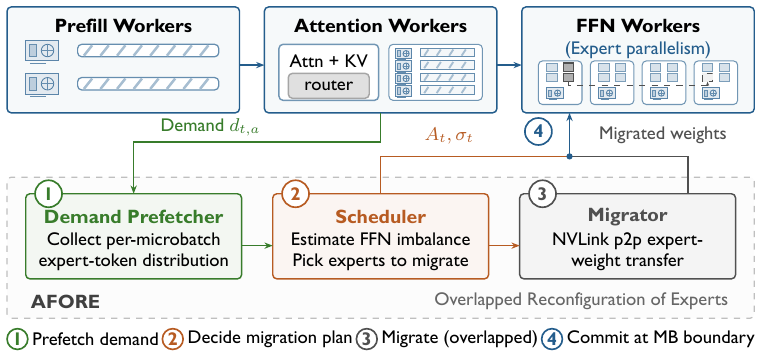}
    \caption{Overall architecture of \sys. \sys adds a layer for overlapped
    reconfiguration of experts between attention and FFN workers to prefetch
    upcoming expert demand, schedule migration-aware placement updates,
    overlap expert migration with in-flight FFN execution, and commit new
    placements at microbatch boundaries.}
    \label{fig:overview}
    \Description{The upper row shows prefill, attention, and FFN workers.
    The lower row shows three functional roles: demand prefetcher,
    scheduler, and migrator. Attention workers count selected expert IDs and
    send demand to the shared CPU scheduler. The scheduler produces replica
    placement and exact work assignment and publishes the target plan for
    FFN workers. Per-worker migrators submit required copies early and ensure
    their completion before the matching microbatch uses the weights. The
    lower boxes denote functions within these workers, not separate worker
    groups.}
\end{figure}

\section{\sys Overview}
\label{sec:overview}

This section presents the overall architecture of \sys. As shown in~\autoref{fig:overview}, \sys is built on top of an AFD-based MoE serving runtime. The system contains three major components: the \textit{demand prefetcher}, the \textit{placement scheduler}, and the \textit{placement migrator}. Together, these components allow \sys to observe upcoming expert demand, determine whether expert placement should be adjusted, and perform expert migration with minimal impact on the AFD pipeline.

\vspace{0.5em}
\noindent\textbf{Overall routine.}
\sys serves MoE inference through separate prefill, attention, and FFN workers. We introduce a layer for overlapped reconfiguration of experts between the attention side and the FFN side, which continuously monitors upcoming expert demand and adapts expert placement before target microbatches start FFN execution. The runtime workflow of \sys proceeds as follows. \underline{First}, when the attention workers process a target microbatch, the demand prefetcher extracts its expert-token distribution before the microbatch reaches the FFN side. \underline{Second}, the placement scheduler uses this prefetched information, together with the current expert placement and FFN worker status, to estimate whether the target microbatch will create FFN-stage imbalance. If reconfiguration is beneficial, the scheduler selects the experts to migrate and determines the target placement. \underline{Third}, the placement migrator executes the selected migration while preceding in-flight microbatches continue computing on FFN workers. \underline{Finally}, once migration completes, \sys commits the new placement at a microbatch boundary, so the target or subsequent microbatches can use the updated expert placement without disrupting in-flight execution.

\section{Scheduling Framework}
\label{sec:sched}

In this section, we present the scheduling framework of \sys. We first formulate
the temporal expert reconfiguration problem under AFD-based MoE serving, and
then introduce the migration-aware scheduling algorithm that determines when and how to adjust expert
placement. \autoref{tab:notation} summarizes the notation
used in this section.

\begin{table}[!t]
\centering\normalsize
\renewcommand{\arraystretch}{1.12}
\caption{Notation in the scheduling problem formulation.}
\label{tab:notation}
\begin{tabularx}{0.80\linewidth}{>{\raggedright\arraybackslash}p{0.18\linewidth} | >{\raggedright\arraybackslash}X}
\hline
\textbf{Symbol} & \textbf{Meaning} \\
\hline
$\mathcal{W},\,N$ & set of FFN expert workers, and their count \\
\hline
$\mathcal{E},\,E$ & set of MoE experts, and their count \\
\hline
$w,\,e$ & a worker index/ an expert index \\
\hline
$m_t$ & target microbatch at scheduling epoch $t$ \\
\hline
$d_t^e$ & token--expert assignments of $m_t$ routed to $e$ \\
\hline
$S_t=\sum_e d_t^e$ & total token--expert assignments in $m_t$ \\
\hline
$P=(x_{e,w})$ & placement; $x_{e,w}{=}1$ iff $e$ has a replica on $w$ \\
\hline
$\mathcal{R}_e(P),\,r_e(P)$ & replica set and replication degree of $e$ \\
\hline
$K_w$ & expert-slot (memory) budget of worker $w$ \\
\hline
$A_t=(a_t^{e,w})$ & integer assignment counts for $m_t$ across workers \\
\hline
$\phi_{e,w}$ & fraction of $e$'s tokens routed to its replica on $w$ \\
\hline
$L_w(A_t)$ & token load of worker $w$ \\
\hline
$M_{\mathrm{tile}}$ & grouped-GEMM M-dimension tile size \\
\hline
$T_w(A_t)$ & grouped-GEMM tile load of worker $w$ \\
\hline
\end{tabularx}
\end{table}

\subsection{Problem Formulation}
\label{subsec:problem-formulation}

We formulate \sys scheduling as an online, microbatch-driven MoE expert placement
problem for an AFD-based MoE serving system. On the FFN side, a set of expert workers
execute in expert parallelism, with each worker hosting a subset of the MoE
experts. \sys proceeds in scheduling epochs. At each epoch, the scheduling algorithm first observes
the expert-token demand of an upcoming target microbatch, and then decides whether to retain the current placement or install a new one by migrating expert replicas across workers.
The objective is to improve
end-to-end AFD performance by relieving the FFN-side bottleneck, while accounting for the overhead that
expert migration introduces.

\vspace{0.5em}
\noindent \textbf{Modeling system execution and expert placement.}
The FFN side comprises $N$ expert workers $\mathcal{W}=\{1,\dots,N\}$ that
jointly host $E$ MoE experts $\mathcal{E}=\{1,\dots,E\}$. Requests are processed
as a stream of microbatches $\dots,m_{t-1},m_t,m_{t+1},\dots$ in execution
order, and a scheduling epoch is associated with each target microbatch $m_t$.
After the attention/routing side finishes $m_t$, the prefetcher
(\S\ref{sec:overview}) exposes its expert-token demand
$d_t=(d_t^1,\dots,d_t^E)$, where $d_t^e\in\mathbb{Z}_{\ge 0}$ is the number of tokens routed to
expert $e$. An expert placement is a binary assignment $P=(x_{e,w})$, where
$x_{e,w}=1$ iff a replica of expert $e$ resides on worker $w$; we write
$\mathcal{R}_e(P)=\{w:x_{e,w}=1\}$ for the replica set of $e$ and
$r_e(P)=|\mathcal{R}_e(P)|$ for its replication degree. A high-load expert may be
placed on multiple workers ($r_e>1$) so that its tokens are served in parallel,
whereas a cold expert needs only a single replica. A placement is feasible iff
every expert has at least one replica and no worker exceeds its device-memory
budget of $K_w$ expert slots:
\begin{equation}
  r_e(P)\ge 1\ \ \forall e\in\mathcal{E},
  \qquad
  \textstyle\sum_{e\in\mathcal{E}} x_{e,w}\le K_w\ \ \forall w\in\mathcal{W}.
  \label{eq:feasible}
\end{equation}

\vspace{0.5em}
\noindent \textbf{FFN load and scheduling objective.}
Let $A_t=(a_t^{e,w})$ denote the work assignment, where
$a_t^{e,w}\in\mathbb{Z}_{\ge 0}$ counts the token--expert assignments executed
by worker $w$ for expert $e$. Each expert's demand is distributed among its
available replicas:
\begin{equation}
  \sum_{w\in\mathcal{W}}a_t^{e,w}=d_t^e\quad\forall e\in\mathcal{E},
  \qquad a_t^{e,w}=0\quad\text{if }x_{e,w}=0.
  \label{eq:assignment-feasible}
\end{equation}
The corresponding routing fraction is $\phi_{e,w}=a_t^{e,w}/d_t^e$ when
$d_t^e>0$, and zero otherwise. The token load of worker $w$ is
\begin{equation}
  L_w(A_t)=\sum_{e\in\mathcal{E}}a_t^{e,w}.
  \label{eq:load}
\end{equation}

Token load alone does not fully describe grouped-GEMM work. The kernel
processes each expert's assignments in M-dimension tiles, and assignments of
different experts cannot share a tile. We define the tile load of worker $w$ as
\begin{equation}
  T_w(A_t)=\sum_{e:a_t^{e,w}>0}
  \left\lceil\frac{a_t^{e,w}}{M_{\mathrm{tile}}}\right\rceil,
  \label{eq:tile-load}
\end{equation}
where $M_{\mathrm{tile}}$ is the M-dimension tile size used by the
grouped-GEMM kernel. Ignoring migration cost, the scheduling objective is to
minimize the maximum worker tile load:
\begin{equation}
  \min_{P,A_t}\ \max_{w\in\mathcal{W}} T_w(A_t),
  \label{eq:obj-tile}
\end{equation}
subject to the placement and assignment constraints
in~\autoref{eq:feasible} and~\autoref{eq:assignment-feasible}. In practice,
changing the placement is not free, so \sys must weigh this benefit
against the cost of migration.

\vspace{0.5em}
\noindent \textbf{Migration cost.}
Switching from the current placement $P_{t-1}$ to a new placement $P_t$ requires
transferring the weights of every replica that $P_t$ adds (removing a replica
requires no weight transfer):
\begin{equation}
  \Delta(P_{t-1}\!\to\!P_t)=\{(e,w): x^{t}_{e,w}=1 \wedge x^{t-1}_{e,w}=0\}.
  \label{eq:delta}
\end{equation}

Transferring these weights incurs a migration cost $M(\Delta)$ (its concrete
realization is in \S\ref{sec:systemdesign}). Because AFD exposes the
target microbatch before it reaches the FFN side, this migration can be
overlapped with the in-flight microbatches still executing on FFN workers; only
the part that exceeds the available overlap window $W_t$ is exposed on the
critical path:
\begin{equation}
  O_t(\Delta)=\max\bigl(0,\ M(\Delta)-W_t\bigr).
  \label{eq:exposed}
\end{equation}
The exposed cost $O_t$, not the raw migration cost, is what a placement change
contributes to the optimization objective.

\vspace{0.5em}
\noindent \textbf{Migration-aware placement problem.}
The load-balancing objective (\autoref{eq:obj-tile}) and the migration cost
(\autoref{eq:exposed}) compete: a new placement can lower the FFN load, but only
by paying to migrate experts. \sys combines them into a single objective.
We estimate the FFN execution time of an assignment as
$F(A_t)=\beta_{\mathrm{tile}}\max_w T_w(A_t)$, where
$\beta_{\mathrm{tile}}$ is the estimated computation time per tile.
Let $A_t^{\mathrm{base}}$ be the assignment for the current target $m_t$
after rebalancing among the existing replicas in $P_{t-1}$.
The net gain of a candidate placement and assignment $(P_t,A_t)$ is
\begin{equation}
  \begin{aligned}
    G_t(P_t,A_t)={}&F(A_t^{\mathrm{base}})-F(A_t) \\
                  &-O_t\bigl(\Delta(P_{t-1}\!\to\!P_t)\bigr).
  \end{aligned}
  \label{eq:gain}
\end{equation}

\noindent At each epoch $t$, \sys selects a placement and assignment that maximize this net gain:
\begin{equation}
  (P_t^\star,A_t^\star)=\arg\max_{P_t,A_t}\ G_t(P_t,A_t),
  \label{eq:p1}
\end{equation}

\noindent subject to~\autoref{eq:feasible} and~\autoref{eq:assignment-feasible}.
The selected pair is installed only when $G_t(P_t^\star,A_t^\star)>0$;
otherwise \sys retains $P_{t-1}$ with assignment $A_t^{\mathrm{base}}$.
This criterion captures the central decision of
\sys: reconfigure only when the FFN time saved by rebalancing outweighs the
migration cost it exposes.

\subsection{Scheduling Algorithm}
\label{subsec:scheduling-algorithm}

Searching the full placement space in~\autoref{eq:p1} for every microbatch
is impractical. \sys therefore fixes each expert's primary placement and
uses the remaining capacity on each worker as \emph{redundant slots}.
Scheduling determines which replicas occupy these slots and how to distribute
work across the resulting placement. Existing replicas can be reused across
microbatches, reducing the need for new weight copies.

The scheduler uses a greedy search in three phases.
\emph{Assignment initialization} adapts the preceding assignment to the target
demand. \emph{Rebalancing within the placement} first redistributes work over
existing replicas without adding weight copies. \emph{Rebalancing beyond the
placement} then explores new replicas and selects a candidate only when its
FFN time savings outweigh the exposed migration cost. This order exploits
the current placement before paying to change it.
The scheduler returns $\Pi_t=(P,A_t,\sigma_t,\Delta_t)$, where $P$ is the
selected placement, $A_t$ gives the assignment counts across workers,
$\sigma_t(e,w)$ identifies expert $e$'s physical slot on worker $w$, and
$\Delta_t$ lists the required new weight copies.
Algorithm~\ref{alg:sched} summarizes the procedure.

\begin{algorithm}[t]
\small
\caption{\sys Per-Microbatch Scheduling}
\label{alg:sched}
\renewcommand{\algorithmicrequire}{\textbf{Input:}}
\renewcommand{\algorithmicensure}{\textbf{Output:}}
\setlength{\baselineskip}{12pt}
\begin{algorithmic}[1]
\Require target microbatch $m_t$, demand $d_t$, projected slot table $C_t$,
protected slots $U_t$, previous assignment $A_{t-1}$, global copy budget $B$,
compactness threshold $\theta\ge 0$, cost estimates
$\beta_{\mathrm{tile}},\,M(\cdot),\,W_t$
\Ensure microbatch plan $\Pi_t$
\Statex $\triangleright$ Stage 1: Assignment initialization
\State $P\leftarrow$ fixed primaries $\cup$ replica bindings recorded in $C_t$;
$\Delta_t\leftarrow\emptyset$; $b\leftarrow 0$
\State $(N_t,T)\leftarrow \operatorname{InitializeAssignment}(A_{t-1},d_t,P;\theta)$
\Statex $\triangleright$ Stage 2: Rebalancing within the placement
\While{$\operatorname{FindWithinMove}(N_t,P,T;\theta)$ returns $(e,h,c,q)$ with $q>\theta$}
  \State move $q$ tiles of $e$ from $h$ to $c$; update $(N_t,T)$
\EndWhile
\Statex $\triangleright$ Stage 3: Rebalancing beyond the placement
\State $F_0\leftarrow\beta_{\mathrm{tile}}\max_w T_w$; $g^\star\leftarrow 0$
\State save $(P,N_t,C_t,\Delta_t)$ as the selected candidate
\While{$b<B$ and $\operatorname{FindBeyondMove}(N_t,P,T,C_t,U_t;\theta)$ returns $(e,h,c,r,q)$ with $q>\theta$}
  \State rebind slot $r$ to $e$ on $c$; update $(C_t,P)$
  \State add $(e,c,r)$ to $\Delta_t$; $b\leftarrow b+1$
  \State move $q$ tiles of $e$ from $h$ to $c$; update $(N_t,T)$
  \State $g\leftarrow F_0-\beta_{\mathrm{tile}}\max_w T_w-O_t(\Delta_t)$
  \If{$g>g^\star$}
    \State save $(P,N_t,C_t,\Delta_t)$ as the selected candidate
    \State $g^\star\leftarrow g$
  \EndIf
\EndWhile
\State restore the selected $(P,N_t,C_t,\Delta_t)$
\State derive $A_t$, $\phi$, and $\sigma_t$ from $N_t$ and projected slots
\State record projected state; return $\Pi_t=(P,A_t,\sigma_t,\Delta_t)$
\end{algorithmic}
\end{algorithm}

\vspace{0.5em}
\noindent\textbf{Assignment initialization.}
The scheduler starts from the fixed primaries and the redundant replicas
recorded in the per-layer projected slot table $C_t$. This table includes
bindings established by earlier plans, even if their copies are still in
flight.

To initialize the work assignment, the scheduler divides expert $e$'s demand
into $k_t^e=\lceil d_t^e/M_{\mathrm{tile}}\rceil$ tiles. The tile assignment
$N_t=(n_t^{e,w})$ records how many tiles of $e$ run on worker $w$, whose
total load is $T_w=\sum_e n_t^{e,w}$.
The scheduler initializes $N_t$ from the preceding assignment $A_{t-1}$,
adapting its per-worker proportions to $k_t^e$ integer tiles while preserving
the total. Without a preceding
assignment, all tiles start on the expert's primary worker. This step adapts
to demand changes while retaining useful assignments from nearby microbatches.

Moving only a few tiles provides limited load relief but can still require
additional routing state at the receiver. The scheduler therefore uses a
nonnegative integer compactness threshold $\theta$, requiring each
rebalancing move to transfer more than $\theta$ tiles. Since experts with
$0<k_t^e\le\theta$ cannot meet this requirement, initialization assigns all
their tiles to the primary. Existing replica bindings remain recorded for
later reuse.

\vspace{0.5em}
\noindent\textbf{Rebalancing within the placement.}
This phase greedily redistributes tiles among replicas already recorded in
$P$. The scheduler scans workers $h$ in decreasing load order and receivers
$c$ in increasing load order. For each pair $(h,c)$, it considers experts
with $n_t^{e,h}>0$ and $c\in\mathcal{R}_e(P)$.
If $T_h-T_c>1$, the scheduler can move
\begin{equation}
  q_e(h,c)=\min\left(n_t^{e,h},
  \left\lfloor\frac{T_h-T_c}{2}\right\rfloor\right)
  \label{eq:tile-move}
\end{equation}
tiles of expert $e$ from $h$ to $c$. The two terms ensure that the move has
enough tiles and does not reverse the pair's load order.
Similarly, only moves with $q_e(h,c)>\theta$ are eligible.
For the first pair admitting such a move, the scheduler
chooses the expert maximizing $q_e(h,c)$, breaking ties by expert index.

The scheduler applies the move, updates worker loads, and repeats the search
until no eligible move remains. Placement $P$
stays fixed, and no weight copy is added to $\Delta_t$. Each move strictly
decreases the load vector, sorted from largest to smallest, in lexicographic
order. This guarantees termination even when several workers share the
maximum load.

\vspace{0.5em}
\noindent\textbf{Rebalancing beyond the placement.}
This phase explores whether new replicas can further reduce the FFN
bottleneck. It also considers only moves with $q_e(h,c)>\theta$.
The search $\operatorname{FindBeyondMove}$ follows the preceding
phase's greedy rule, but requires $c\notin\mathcal{R}_e(P)$ and a safe
redundant slot $r$ on $c$. Each move updates the current candidate plan by
binding slot $r$ to expert $e$ and assigning $q_e(h,c)$ tiles to the replica.
This updates $C_t$ and $P$, replaces any previous binding in $r$, and adds
$(e,c,r)$ to $\Delta_t$.

The scheduler evaluates each updated plan using the net gain
in~\autoref{eq:gain}. The preceding phase produces the baseline assignment
$A_t^{\mathrm{base}}$, giving a fixed FFN time $F_0=F(A_t^{\mathrm{base}})$. After each move,
the scheduler subtracts the candidate's FFN time and exposed migration cost
from $F_0$. This calculation uses the supplied estimates
$\beta_{\mathrm{tile}}$, $M(\cdot)$, and $W_t$, accounting for the exposed
cost of the entire accumulated copy list $\Delta_t$.
Throughout the search, it saves the candidate with the highest positive net
gain seen so far. If none is found, it retains the preceding phase's plan.
Search can continue
through candidates with nonpositive gain, since several moves may be needed
to reduce equally loaded bottleneck workers.

Each candidate must satisfy the slot and copy constraints. A global copy
budget $B$ limits the total number of new replicas for each target microbatch.
The counter $b$ increases by one for each planned copy to a destination worker.
For each receiver, the scheduler first chooses an empty
redundant slot; otherwise, it chooses the least recently used slot whose
replica has no work in $N_t$. Slots used by the preceding microbatch
($U_t$) or reserved by an earlier move in this phase cannot be rebound.
If no safe slot exists, the receiver is skipped.

The search ends when $b=B$ or no feasible move exceeds the tile threshold.
The scheduler then restores the saved placement,
assignment, slot bindings, and copy list. All changes during search are
tentative: only the selected candidate is recorded and passed to execution.

The scheduler fills the selected tile assignment with tokens, leaving at most
one partially filled tile per expert, and returns the resulting plan.
It retains the selected assignment and slot bindings for the next microbatch.

\section{\sys Design and Implementation}
\label{sec:systemdesign}

Section~\ref{sec:sched} constructs a microbatch plan $\Pi_t$ for each target
microbatch. This section addresses two questions left by that algorithm: how
can the target's routing demand reach the scheduler and the
resulting plan reach FFN workers in time, and how can those workers prepare and
apply the plan while earlier microbatches are still executing? As shown in
Figure~\ref{fig:overview}, \sys prefetches demand from the attention side,
publishes each plan through a shared-memory queue, and overlaps replica
migration with preceding FFN computation. We describe these mechanisms in
turn.

\Needspace{6.3in}
\subsection{Demand Prefetching and Plan Delivery}
\label{subsec:demand-plan-delivery}

\begin{wrapfigure}{r}{3.5in}
    \centering
    \includegraphics[trim=0bp 2.36bp 0bp 0bp,clip,width=\linewidth]{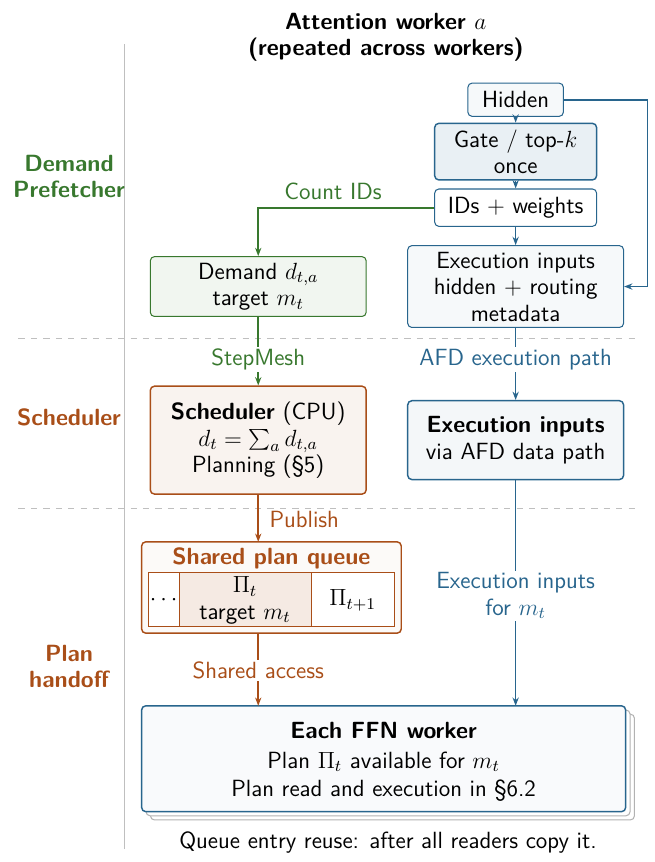}
    \caption{Demand prefetching and plan delivery for $m_t$,
    expanding Steps 1--2 in Figure~\ref{fig:overview}.
    Attention forms routing demand and execution inputs; the CPU scheduler
    aggregates demand and prepares the plan while inputs follow the AFD path.
    The shared-memory queue makes the published plan available to all FFN
    workers. Queue entries are reused after all workers copy them.}
    \label{fig:demand-plan-control}
    \Description{Three horizontal bands show the demand and plan control path
    and its interfaces with the existing AFD runtime. First, one representative
    attention worker applies gate and top-k once. The selected expert IDs are
    counted to produce a demand vector, while the hidden states, expert IDs,
    and routing weights form a separate execution payload. Second, StepMesh
    delivers demand vectors from all attention workers to the CPU
    scheduler, which aggregates them and invokes the scheduling algorithm
    in Section 5. The execution payload independently follows the AFD
    execution path. Third, the scheduler publishes the plan
    through a shared-memory queue, making the same target-tagged plan
    available to every FFN worker. Entries can be reused only after every
    reader has copied them; plan consumption is shown in
    Section~\ref{subsec:overlapped-plan-execution}.}
\end{wrapfigure}

In a conventional expert-parallel MoE runtime, routing information becomes
available only when the FFN workers are about to execute the experts, leaving
little time for reconfiguration. AFD provides an earlier opportunity: the
attention side produces the hidden states before the corresponding FFN
execution. \sys uses this interval to obtain the target microbatch's routing
demand and initiate placement scheduling, as illustrated
in~\autoref{fig:demand-plan-control}.

\vspace{0.5em}
\noindent\textbf{Extracting routing demand.}
After producing the hidden states for a MoE layer, each attention worker runs
the model's gate and top-$k$ to obtain the selected expert IDs and routing
weights. It counts the IDs to form $d_{t,a}$, where $d_{t,a}^e$ is the number
of token--expert assignments to expert $e$ in that worker's portion of the
target microbatch. The IDs and weights are also reused for FFN execution,
avoiding a second gate and top-$k$ computation.

\vspace{0.5em}
\noindent\textbf{A separate path for demand.}
To deliver the newly formed $d_{t,a}$, StepMesh~\cite{wang2025step} writes
each compact vector directly into a registered host-memory mailbox on the
FFN node. The hidden states, expert IDs, and routing weights travel along
another path to the FFN workers for execution. This separation makes demand
available to the CPU scheduler before the full execution input is assembled,
leaving more of the AFD lookahead window for scheduling and subsequent expert
migration. Each vector is associated with its serving step, layer,
microbatch, and source attention worker, allowing the scheduler to collect
demand in target order.

\vspace{0.5em}
\WFclear
\noindent\textbf{Shared scheduling and plan delivery.}
Once vectors from all attention workers have arrived for a target microbatch,
the shared CPU scheduler forms $d_t=\sum_a d_{t,a}$ and generates $\Pi_t$
using the algorithm in Section~\ref{subsec:scheduling-algorithm}. It publishes
each target-tagged plan once to a shared-memory ring. Because all FFN workers
map this ring, one publication makes the same plan available to every worker
without separate plan messages. To remain bounded while protecting slow readers,
the queue reuses a slot only after all FFN workers have copied its plan.
Plan consumption and execution are described in
Section~\ref{subsec:overlapped-plan-execution}.

\begin{figure}[t]
    \centering
    \includegraphics[trim=0bp 4.58bp 0bp 0bp,clip,width=6.3in]{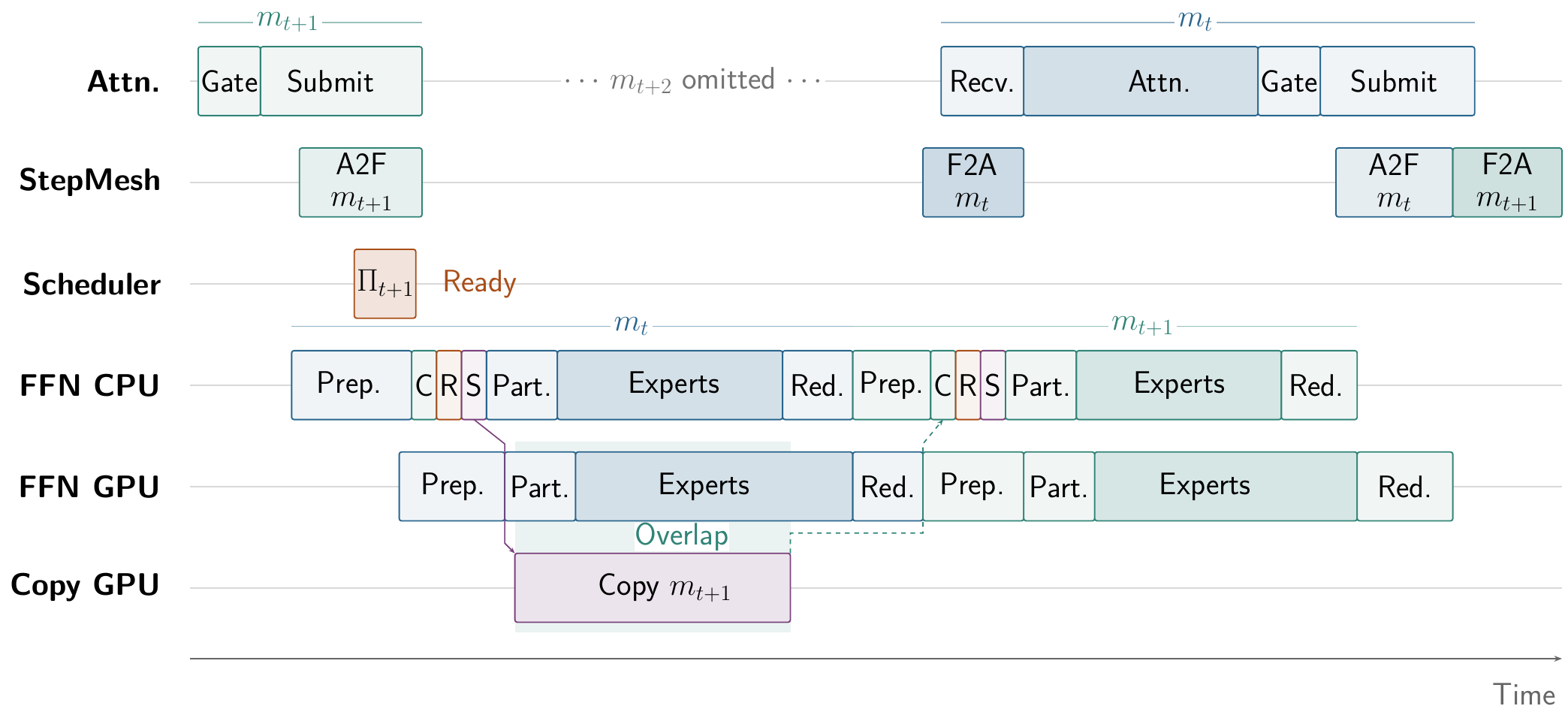}
    \caption{Overlapped migration and plan execution for FFN microbatches
    $m_t$ and $m_{t+1}$, expanding Steps 3--4 in
    Figure~\ref{fig:overview}. Shading covers the copy for $m_{t+1}$ during
    $m_t$'s FFN execution. The first C checks $m_t$'s replicas, R reads
    $\Pi_{t+1}$, and S submits its copies. The solid arrow links S to GPU
    copy execution; the dashed arrow links copy completion to $m_{t+1}$'s C,
    which enqueues a model-stream wait if the copy remains unfinished. The
    later R/S begins preparation for the following microbatch.
    A2F/F2A denote StepMesh transfers.}
    \label{fig:placement}
    \Description{Six timeline rows show an excerpt of a three-microbatch
    pipeline. The FFN CPU and GPU stages of both $m_t$ and $m_{t+1}$ in layer
    $\ell$ show input preparation, partition, expert computation, and reduction.
    Attention shows $m_{t+1}$'s gate
    and send tail, an explicitly omitted interval for $m_{t+2}$, and $m_t$ in the next layer.
    StepMesh displays the corresponding A2F and F2A transfers. The CPU
    scheduler prepares the plan for $m_{t+1}$. The FFN CPU checks $m_t$'s
    replicas, reads the next plan, and submits copies while continuing
    $m_t$'s execution. A solid arrow connects S to the copy-stream block,
    which overlaps $m_t$'s FFN work. A dashed arrow
    connects copy completion to the next C readiness check; pending copies
    induce a model-stream wait before $m_{t+1}$ consumes the weights.
    The following R/S starts preparation for the subsequent microbatch.
    Unshown work remains part of the three-microbatch pipeline.}
\end{figure}

\subsection{Placement Migrator and Plan Execution}
\label{subsec:overlapped-plan-execution}

After the scheduler publishes $\Pi_{t+1}$ for $m_{t+1}$ to the shared ring,
FFN workers must prepare its required replicas before the target can use the
plan. This work takes place while $m_t$ is still in flight: weights it uses
must remain valid, and new replicas cannot be consumed until their copies
complete. Figure~\ref{fig:placement} shows the handoff across the FFN model
thread, model stream, and migration stream.

\vspace{0.5em}
\noindent\textbf{Plan read and copy submission.}
Before partitioning $m_t$'s token--expert pairs, each FFN worker reads
$\Pi_{t+1}$ from the shared ring (R in Figure~\ref{fig:placement}) and
validates its target and sequence to reject a stale or out-of-order plan.
It immediately submits the missing copies in $\Delta_{t+1}$ (S) to overlap
migration with the remainder of $m_t$'s execution, while keeping $\Pi_{t+1}$
pending until $m_{t+1}$ arrives. Replicas already resident in their slots
need no further copy.

\vspace{0.5em}
\noindent\textbf{Overlapped migration.}
To run the copies alongside $m_t$'s FFN computation, the migrator batches
these GPU-to-GPU P2P transfers from fixed primary owners to reserved slots
on a dedicated migration stream. The source primary may differ from the
worker that relinquishes the tiles.
Meanwhile, the model thread can continue $m_t$'s partition, expert, and
reduce/reply stages and prepare later inputs, with their GPU work running on
the model stream. To avoid overwriting active replicas, the scheduler excludes
slots still used by $m_t$ from the copy destinations. A completion event
marks when the copy batch finishes, allowing the target to check replica
readiness at C.

\vspace{0.5em}
\Needspace{3\baselineskip}
\noindent\textbf{Readiness and exact partitioning.}
When $m_{t+1}$ reaches the FFN stage, workers first all-gather its hidden
states and routing metadata received through StepMesh. This gives every
worker the same input for exact partitioning. Each worker then retrieves
the pending $\Pi_{t+1}$ and checks the required replicas' copy-completion
events (C in Figure~\ref{fig:placement}). Completed copies require no wait;
for any still in flight, the worker enqueues a model-stream wait before using
the new replica. With the weights ready, workers
partition the target's token--expert pairs according to $A_{t+1}$ and
$\sigma_{t+1}$.
The scheduler's integer count $a_{t+1}^{e,w}$ specifies how many of expert
$e$'s pairs execute on worker $w$. Because all FFN workers share the gathered
routing result, they can number each occurrence of $e$ in the same order and
select disjoint intervals without another token transfer. For selected redundant
workers $w_1,\ldots,w_{k_e}$ in rank order, the cumulative boundaries are
\begin{equation}
  o_0=0,\qquad o_j=o_{j-1}+a_{t+1}^{e,w_j}.
  \label{eq:partition-offsets}
\end{equation}
Worker $w_j$ executes ordinals in $[o_{j-1},o_j)$, and the fixed primary owner
executes the remainder $[o_{k_e},d_{t+1}^e)$. These disjoint intervals realize
$A_{t+1}$ exactly. Each worker masks pairs outside its interval and maps retained
logical expert IDs to fixed primary slots or redundant slots specified by
$\sigma_{t+1}$, preserving the original routing weights. The model all-reduce
combines partial outputs before StepMesh returns the corresponding slices to
attention.

\section{System Evaluation}
\label{sec:eval}

We evaluate the design and implementation of \sys by answering the following research questions:

\begin{itemize}[leftmargin=*]
  \item \textbf{(RQ1)} How does \sys compare against state-of-the-art baselines in output throughput and P95 inter-token latency?
   \vspace{0.5em}
  \item \textbf{(RQ2)} How efficient and effective is \sys's scheduling algorithm in practice?
   \vspace{0.5em}
  \item \textbf{(RQ3)} How much does each system design contribute to \sys's performance gains?
   \vspace{0.5em}
  \item \textbf{(RQ4)} How effectively does AFD pipeline overlap hide expert-migration overhead?
\end{itemize}

\subsection{Experimental Setup}
\label{subsec:setup}

\noindent\textbf{Hardware environment.}
We conduct all experiments on a cluster of two machines, each equipped with
eight NVIDIA A100 GPUs. Within each node, the eight GPUs are interconnected through NVLink
and NVSwitch, providing a peak unidirectional bandwidth of
300~GB/s per GPU. For inter-node communication, each node provides eight RDMA-enabled bonded network interfaces
with 50~GB/s of unidirectional bandwidth per bond.

\vspace{0.5em}
\noindent\textbf{Models and workloads.}
We evaluate \sys on a 110B-parameter MoE model (i.e., \texttt{GLM-4.5-Air}),
served under a PD+AFD deployment. We construct four representative serving
workloads from ShareGPT~\cite{zheng2023judging},
FineWeb~\cite{penedo2024fineweb},
CodeForces~\cite{penedo2025codeforces}, and
GSM8K~\cite{cobbe2021gsm8k}. These workloads are diverse, covering
conversational requests, document-style inputs, code-generation tasks, and
mathematical-reasoning queries, respectively. We replay the traces to generate
dynamic serving workloads whose expert-token distributions shift over time.

\vspace{0.5em}
\noindent\textbf{Evaluation metrics.}
We report standard LLM serving metrics. We measure end-to-end output
throughput in tokens per second and P95 inter-token latency (ITL), which
captures the tail latency of decode-token generation. For migration-specific
analysis, we report raw migration cost and exposed migration cost on the
target microbatch's critical path.

\vspace{0.5em}
\noindent\textbf{Baselines.}
We compare \sys against representative expert-placement and
reconfiguration policies. For a fair comparison, we implement all baselines
inside the same AFD serving runtime and run them with the same model, hardware,
GPU allocation, batching policy, and workloads:
\begin{itemize}[leftmargin=*]
\item \textbf{Static}, a fixed-placement baseline that evenly assigns experts
to FFN workers at initialization and never changes the placement during serving.
 \vspace{0.5em}
\item \textbf{EPLB}~\cite{deepseek2025expert}, a statistics-based
reconfiguration policy that periodically rebalances expert placement using
load statistics collected over a past profiling window.
 \vspace{0.5em}
\item \textbf{HarMoEny}~\cite{doucet2025harmoeny}, a reactive load-balancing
policy that redistributes tokens and creates redundant expert replicas based on
the routing demand observed at runtime.
 \vspace{0.5em}
\item \textbf{Lina}~\cite{li2023accelerating}, a prediction-based
reconfiguration policy that predicts upcoming hot experts from recent expert
selection patterns and replicates them in advance.
 \vspace{0.5em}
\item \textbf{Libra}~\cite{yang2026libra}, a prediction-based
reconfiguration policy that prefetches expert replicas based on predicted
next-layer top-$k$ activations and redistributes tokens according to
the actual routing demand.
\end{itemize}

We further evaluate ablated versions of \sys in~\S\ref{subsec:ablation-study}
to isolate the contributions of target-demand visibility and
pipeline-overlapped migration.

\begin{figure}[t]
    \centering
    \includegraphics[trim=0bp 12.37bp 0bp 0bp,clip,width=6.3in]{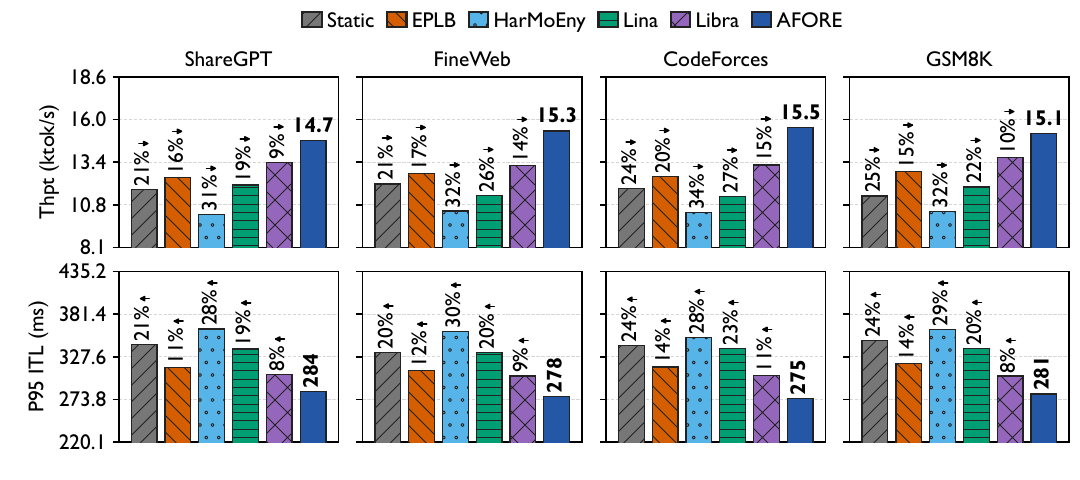}
    \caption{\textbf{\underline{Upper:}} Output throughput across four
    workloads. \textbf{\underline{Lower:}} P95 inter-token latency under the
    same workloads.}
    \label{fig:e2e-results}
\end{figure}

\Needspace{4\baselineskip}
\subsection{End-to-end Performance \textbf{(RQ1)}}
\label{subsec:e2e}

\autoref{fig:e2e-results} compares \sys with the baselines across the four
workloads. \sys achieves the highest output throughput and the lowest
P95 ITL on all workloads. Compared with the strongest competing baseline
for each workload, \sys improves throughput by 10.1--17.6\% and
reduces P95 ITL by 7.1--9.5\%.

The improvement comes from two effects. \underline{First}, temporal reconfiguration reduces
FFN-side stragglers by adapting expert placement to the routing demand that is
about to enter the FFN stage. \underline{Second}, \sys overlaps
expert migration with the AFD pipeline, avoiding the critical-path migration
cost that limits more reactive policies. As a result, \sys improves both
throughput and tail token latency.



The gains persist across workloads, i.e., conversational (\texttt{ShareGPT}), document-style (\texttt{FineWeb}),
coding (\texttt{CodeForces}), and mathematical-reasoning (\texttt{GSM8K}) workloads. Despite
their different request characteristics and routing patterns, \sys consistently
improves the MoE serving efficiency, showing that temporal expert reconfiguration is
robust across diverse workload types.

\Needspace{6\baselineskip}
\subsection{Scheduler Overhead and Solution Quality \textbf{(RQ2)}}
\label{subsec:scheduling-efficiency}

\noindent\textbf{Evaluation configuration.}
We evaluate whether \sys's scheduler is efficient enough to run at
microbatch granularity while producing high-quality plans. We replay
routing traces from the four workloads under EP degrees 2, 4, 8, 16,
32, 64, and 128, sampling the same 200 steady-state targets
(50 per workload) at each EP degree. For scheduling overhead, we report
the median scheduler core latency across the sampled targets. For solution
quality, we compare \sys's maximum worker tile load with the optimum
under the same placement and copy constraints.
Optimality is certified by matching the average tile-load lower bound or by
exact optimization under the same constraints. We report the relative gap
between \sys's maximum worker tile load and the optimum.

\begin{figure}[t]
    \centering
    \includegraphics[trim=0bp 8.70bp 0bp 0bp,clip,width=4.5in]{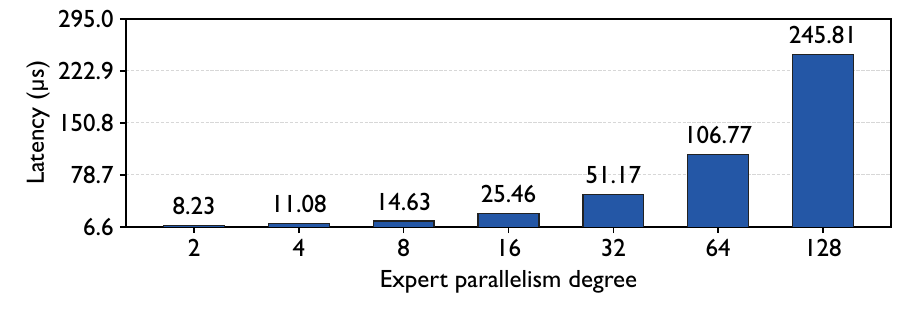}
    \caption{Median scheduler core latency of \sys across EP degrees.}
    \Description{Seven blue bars show median scheduler core latency increasing
    from 8.23 microseconds at EP 2 to 245.81 microseconds at EP 128.}
    \label{fig:scheduler-overhead}
\end{figure}

\begin{table}[!t]
    \centering
    \caption{Scheduler solution quality. Gaps are relative to the optimum.}
    \label{tab:scheduler-summary}
    \normalsize
    \renewcommand{\arraystretch}{1.12}
    \begin{tabularx}{4.5in}{@{}l | *{6}{>{\centering\arraybackslash}X |} >{\centering\arraybackslash}X@{}}
\hline
\textbf{EP} & \textbf{2} & \textbf{4} & \textbf{8} & \textbf{16} & \textbf{32} & \textbf{64} & \textbf{128} \\
\hline
Optimal / 200 & 196 & 197 & 200 & 200 & 200 & 200 & 200 \\
\hline
Mean gap (\%) & 0.04 & 0.03 & 0 & 0 & 0 & 0 & 0 \\
\hline
Max gap (\%) & 3.60 & 1.89 & 0 & 0 & 0 & 0 & 0 \\
\hline
\end{tabularx}\hspace*{\fontdimen2\font}
\end{table}
\FloatBarrier

\vspace{0.5em}
\Needspace{5\baselineskip}
\noindent\textbf{Scheduling overhead.}
As shown in~\autoref{fig:scheduler-overhead}, \sys's scheduler adds only
sub-millisecond planning overhead across all tested EP degrees. Its median
latency is 8.23\,$\mu$s at EP=2 and remains 245.81\,$\mu$s even at
EP=128. This low planning cost makes the scheduler suitable for
microbatch-level reconfiguration as the number of workers increases.

\vspace{0.5em}
\noindent\textbf{Solution quality.}
\autoref{tab:scheduler-summary} compares \sys's greedy scheduler with
the certified optimum in maximum worker tile load. At EP=2 and EP=4,
the mean optimality gap is at most 0.04\%, with a worst-case gap of
3.60\% across the two settings. At every tested EP degree from 8 to 128,
\sys achieves the optimum on all 200 targets.
Overall, it achieves the optimal maximum worker tile load
on 99.5\% of the 1,400 targets.

These results show that \sys's scheduler provides an effective trade-off
for online serving: it combines the low latency needed for frequent
reconfiguration with near-optimal load balance, even at high EP degrees.

\subsection{Component-wise Ablation \textbf{(RQ3)}}
\label{subsec:ablation-study}

\noindent\textbf{Evaluation configuration.}
We isolate the effects of demand prefetching and migration overlap.
\emph{Full \sys} enables both mechanisms.
\emph{w/o Prefetch} replaces the target microbatch's demand with the
previous step's demand at the same layer and microbatch position, while
retaining the same scheduler and migration mechanism.
\emph{w/o Overlap} retains demand prefetching and the same scheduler,
but places expert migration on the critical path. \emph{Static} disables runtime reconfiguration
and serves as a fixed-placement reference. All configurations use the
same model, workload, resource allocation, and batching policy.

\begin{figure}[!t]
    \centering
    \includegraphics[trim=0bp 12.37bp 0bp 0bp,clip,width=6.3in]{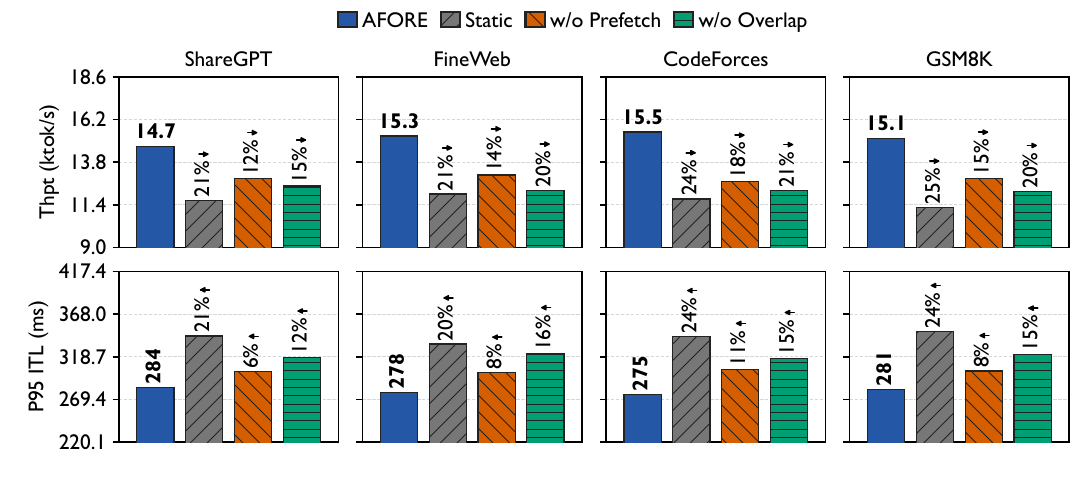}
    \caption{\textbf{\underline{Upper:}} Output throughput for the
    component-wise ablation across four workloads.
    \textbf{\underline{Lower:}} P95 inter-token latency under the same
    configurations.}
    \label{fig:ablation}
\end{figure}

\vspace{0.5em}
\Needspace{4\baselineskip}
\noindent\textbf{End-to-end contribution.}
\autoref{fig:ablation} reports output throughput and P95 inter-token
latency across the four workloads. Compared with Static, full \sys
improves throughput by 26.0--34.1\% (29.8\% on average) and
reduces P95 ITL by 16.7--19.6\% (18.2\% on average).
The combined design consistently improves serving performance over
fixed expert placement.

\vspace{0.5em}
\noindent\textbf{Demand prefetching.}
Disabling demand prefetching reduces throughput by 12.4--17.9\% and
increases P95 ITL by 6.3--10.5\% relative to full \sys.
Without the target microbatch's routing demand, the scheduler acts on
historical observations and may optimize for a load distribution that
has already changed. The degradation across workloads illustrates the
value of timely routing information for expert reconfiguration.

\begin{table}[t]
    \centering
    \normalsize
    \renewcommand{\arraystretch}{1.12}
    \caption{Ablation study configurations.}
    \label{tab:ablation-config}
    \begin{tabularx}{4.5in}{@{}>{\raggedright\arraybackslash}p{0.18\linewidth} | *{2}{>{\centering\arraybackslash}X |} >{\centering\arraybackslash}X@{}}
        \hline
        \textbf{Configuration} & \textbf{Reconfig.} & \textbf{Prefetch} & \textbf{Overlap} \\
        \hline
        Static        & -- & -- & -- \\
        \hline
        w/o Prefetch  & \checkmark & -- & \checkmark \\
        \hline
        w/o Overlap   & \checkmark & \checkmark & -- \\
        \hline
        \sys           & \checkmark & \checkmark & \checkmark \\
        \hline
    \end{tabularx}
\end{table}
\FloatBarrier

\FloatBarrier

\vspace{0.5em}
\noindent\textbf{Migration overlap.}
Without pipeline overlap, throughput drops by 15.2--21.3\%, while
P95 ITL rises by 12.2--16.0\%. Delaying expert copies exposes their
latency on the target microbatch's critical path, even though the
scheduler still uses its actual routing demand.

Therefore, demand prefetching and migration overlap play complementary
roles: prefetching improves \emph{which} placement is selected, while
overlap reduces the cost of \emph{installing} that placement.

\Needspace{5\baselineskip}
\subsection{Migration Hiding Effectiveness \textbf{(RQ4)}}
\label{subsec:breakdown}

\noindent\textbf{Evaluation configuration.}
We measure the raw migration cost from copy submission to completion
and the exposed migration cost as the remaining wait for copies when
the target microbatch needs them. Their difference is the migration
time hidden by pipeline overlap. We also measure the lookahead window
from copy submission to the target microbatch's first use of the replicas.

\begin{figure}[t]
    \centering
    \includegraphics[trim=0bp 10.30bp 0bp 0bp,clip,width=5.4in]{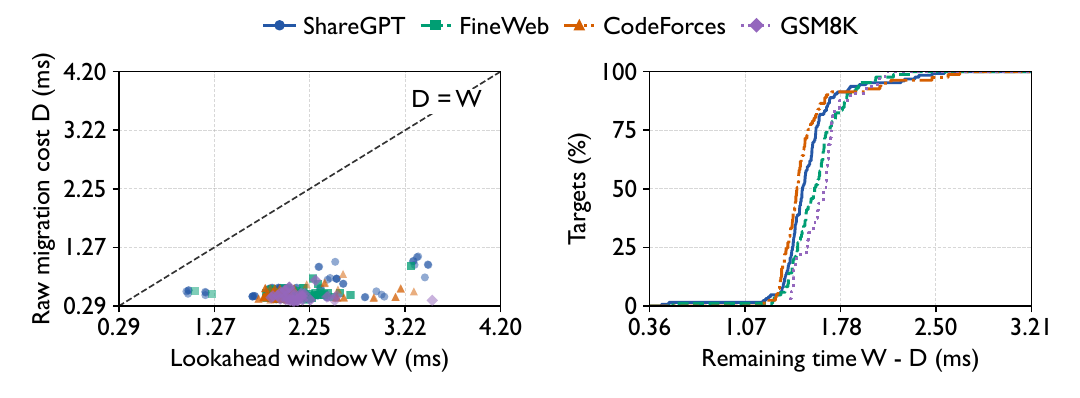}
    \caption{\textbf{\underline{Left:}} Raw migration cost $D$ versus
    lookahead window $W$.
    \textbf{\underline{Right:}} CDF of time remaining before use
    (minimum across ranks per target).}
    \label{fig:runtime-breakdown}
\end{figure}

\vspace{0.5em}
\noindent\textbf{Migration cost.}
The left panel of~\autoref{fig:runtime-breakdown} compares raw migration
cost with the lookahead window across the four workloads. Mean raw
migration cost ranges from 0.453--0.495\,ms, compared with a
mean lookahead window of 2.008--2.119\,ms. All 324 sampled
targets complete their required copies before use, yielding zero exposed
migration cost. Thus, pipeline overlap fully hides the migration time
in these samples and keeps it off the target microbatch's critical path.

\vspace{0.5em}
\noindent\textbf{Time remaining before use.}
The right panel of~\autoref{fig:runtime-breakdown} reports the time
between copy completion and first use, taking the minimum across
participating ranks for each target. The median remaining time ranges
from 1.459--1.668\,ms across workloads, with a minimum of
0.447\,ms. This remaining window allows additional migration
delay to be absorbed before it becomes exposed on the target
microbatch's critical path.

\section{Conclusion}
\label{sec:conclusion}

In this paper, we presented \sys, a timely expert reconfiguration system for attention-FFN disaggregated MoE serving. \sys addresses the FFN-side load imbalance amplified by AFD by exploiting two properties of the disaggregated pipeline: upcoming microbatch demand is visible before FFN execution, and expert migration can make progress while preceding microbatches are still executing. Based on these key observations, \sys formulates expert reconfiguration as a microbatch-aware scheduling problem that jointly considers near-future expert demand, FFN-stage bottlenecks, and exposed migration overhead. \sys's scheduler uses a primary-plus-redundant expert layout, a migration-aware cost model, and greedy load redistribution to decide when to install redundant expert replicas. At runtime, attention workers send compact vectors derived from the target routing directly to the scheduler. \sys then installs missing replicas with GPU-to-GPU transfer and applies routing updates at microbatch boundaries without disrupting in-flight execution.
Our evaluation shows that timely expert reconfiguration improves end-to-end serving efficiency across diverse workloads. Compared with the strongest baseline, \sys improves output throughput by 10.1--17.6\% and reduces P95 inter-token latency by 7.1--9.5\%. The scheduler remains lightweight enough for microbatch-level operation, with sub-millisecond latency even at high expert-parallelism degrees. Component-wise ablation isolates the benefits of true target-microbatch demand and early copy submission. Migration profiling then separates raw migration cost from exposed migration cost and quantifies how often the AFD lookahead window hides replica installation. These results demonstrate that AFD exposes not only a new bottleneck for MoE serving, but also the temporal structure needed to mitigate it efficiently.

\FloatBarrier
\Needspace{6\baselineskip}
\bibliographystyle{plainnat}
\bibliography{ref}
\end{document}